\documentclass{sig-alternate}

\usepackage{epstopdf}
\usepackage{algorithm}
\usepackage{algpseudocode}
\usepackage{multirow}
\usepackage{color}
\newtheorem{theorem}{Definition}
\begin{document}


\setcopyright{acmcopyright}

\doi{xx.yyy/zzz_d}

\isbn{123-4567-24-567/08/06}
\acmPrice{\$15.00}

\conferenceinfo{WISCS}{'16 Vienna, Austria}

\acmPrice{\$15.00}

%
\conferenceinfo{WISCS '16}{October 24, 2016, Vienna, Austria}

\title{Shall we collaborate? A model to analyse the benefits of information sharing}

%
%
%
%
%

\numberofauthors{4} 
%
\author{
%
%
%
\alignauthor
Roberto Garrido-Pelaz\\
       \affaddr{Computer Security Lab}\\
       \affaddr{Carlos III University of Madrid}\\
       \affaddr{Madrid, Spain}\\
       \email{rgarrido@pa.uc3m.es}
\alignauthor
Lorena Gonz\'{a}lez-Manzano\\
       \affaddr{Computer Security Lab}\\
       \affaddr{Carlos III University of Madrid}\\
       \affaddr{Madrid, Spain}\\
       \email{lgmanzan@inf.uc3m.es}
\and
\alignauthor
Sergio Pastrana\\
       \affaddr{Computer Security Lab}\\
       \affaddr{Carlos III University of Madrid}\\
       \affaddr{Madrid, Spain}\\
       \email{spastran@inf.uc3m.es}
}

\maketitle
\begin{abstract}
Nowadays, both the amount of cyberattacks and their sophistication have considerably increased, and their prevention is of concern of most of organizations. Cooperation by means of information sharing is a promising strategy to address this problem, but unfortunately it poses many challenges. Indeed, looking for a win-win environment is not straightforward and organizations are not properly motivated to share information. This work presents a model to analyse the benefits and drawbacks of information sharing among organizations that presents a certain level of dependency. The proposed model applies functional dependency network analysis to emulate attacks propagation and game theory for information sharing management. We present a simulation framework implementing the model that allows for testing different sharing strategies under several network and attack settings. Experiments using simulated environments show how the proposed model provides insights on which conditions and scenarios are beneficial for information sharing. 
\end{abstract}


%
%

%
%



\section{Introduction} \label{sec:intro}

In the last decade, cyber attacks have considerably increased and nowadays cyber-crime is considered a stable and growing industry \cite{karsberg2015report,ponemon2016costs,subramanian2014study}. Cybersecurity prevention, detection and response is an ongoing challenge that needs constant and new efforts to protect critical infrastructures, organizations, enterprises and individual welfare. After an intrusion or attack has succeeded, it is important to perform an incident investigation to determine causes and consequences, and to update the security measures that failed (e.g. developing new IDS rules, updating blacklists, etc.). Information gathered in this process is a valuable asset, but unfortunately it is usually kept secret in the inner boundaries of the companies, organizations or even national governments.

Cooperation between different parties has emerged as an essential strategy to improve cybersecurity prevention. National and international efforts encourage the application of cooperation-based solutions to address cybersecurity problems. For example, the US National Security Presidential Directive 54 and Homeland Security Presidential Directive 23 \cite{eeuu2008nspd54} was launched in 2008 to get effective cybersecurity environments. European Parliament has also agreed a Network and Information Security (NIS) Directive \cite{eu2016nisdirective} that will enter into force in August 2016. The benefits of cooperation are even higher in case of critical infrastructures, which highly rely on Information and Communication Technologies (ICT) that also share services among them. The development of reliable and resilience infrastructures should be viewed as an overall strategy rather than a single, independent task. Sharing information related to cybersecurity can be helpful in this regard.

However, there are many challenges and drawbacks that discourage organizations to share information \cite{petrenj2013information}. First, it is critical to look for a win-win environment in which all entities are benefited and where free-riders (i.e. those entities that benefit from others but do not cooperate) are avoided. Second, the reputation of targeted entities is an asset to protect, and one of the main drawbacks of sharing information is precisely the loss of privacy. Third, it is important to take into account some form of trust management, where companies can trust each other to incentive sharing \cite{Luiijf2015}.

Information sharing facilitates a common understanding of threats and thus, it benefits organizations in aspects like quality of risk management, incident response or recovery management. However, despite the clear benefits, neither private nor public organizations are prone to collaborate unless there are tangible incentives that motivate them to do it. Recent works have identified information sharing as a cost-benefit Prisoner's dilemma solved using game theory \cite{tosh2015evolutionary,Luiijf2015}. However, these proposals focus on the mathematical analysis of games between two players, without taking into account the overall network of entities and their functional dependencies. Moreover, these works do not consider how cyberattacks affect other entities in the network due to their propagation.

In this work, we present a model for cybersecurity information sharing among dependent organizations being impacted by different cyberattacks. The model allows simulating real networks and different adversarial capabilities by establishing different attack patterns, assets and dependencies between partners. It applies a propagation algorithm to infer how the entire network is affected by independent cyberattacks, and simulates different sharing strategies. Then, it outputs analytical results that may help security staff to determine under which circumstances it is interesting to share or not. The proposed model does not aim at providing the ground truth about sharing or not, but it helps organizations and governments to take this decision by simulation. 
In this work we present the following contributions:
\begin{enumerate}
\item We describe a model that considers the propagation of the impact of cyberattacks on a network and that applies different strategies for information sharing to mitigate such impact. The model applies Functional Dependency Network Analysis (FDNA) for attacks propagation and game theory for information sharing management. 
\item We have developed a publicly available simulation framework that implements the model. This framework allows to simulate and study test cases by analysing results from both network point of view and particular nodes.
\item Using the simulation framework, we have applied the model in different scenarios having different adversarial settings. Our experimental work shows how the model can provide knowledge about which sharing strategies are better under different network conditions and attack patterns.
\end{enumerate}

The rest of the paper is organized as follows. Section \ref{1sec:related-work} reviews the literature. Section \ref{1sec:background} presents some background on functional dependency network analysis and game theory. Then the model overview and description are presented in Section \ref{1sec:model}, and Section \ref{1sec:experimentation} details the implemented framework. Finally, Section \ref{1sec:conclusions} presents the conclusions and ongoing work.


\section{Related work}\label{1sec:related-work}




Several works have proposed the use of game theory to analyse the trade-off in terms of incentives and costs of sharing information among entities. Naghizadeh et Liu \cite{naghizadeh2016inter} propose folk theorems and use an analytical method to study how the role of private and public monitoring through inter-temporal incentives can support degree of cooperation. Similar to this work, we also propose a game theory based model where utilities of sharing information are calculated upon gains and costs. However, instead of using historical public available actions, we propose immunization factor and reputation as the main variables for incentive sharing. Furthermore, the work in \cite{naghizadeh2016inter} identifies the cost of disclosure as one of the main drawbacks related with  information sharing. While we also consider privacy and disclosure costs as two key drawbacks for sharing, we also introduce a third variable that affects costs, i.e. trust.

Tosh et al. \cite{tosh2015evolutionary} use game theory to help organizations to decide whether to  share information or not, using the CYBEX framework \cite{rutkowski2010cybex}. Authors use evolutionary game theory in order to attain an evolutionary stable strategy (ESS) under various conditions. These conditions are extracted through simulation with synthetic data in a non-cooperative scenario with rational and profit-seeking firms. The main incentive for sharing is the information received, and thus the knowledge gained. In our work we also consider this knowledge as an incentive for sharing. 

Khouzani et al. \cite{khouzani2014strategic} present a two stage Bayesian game between two firms  to help to decide how much to invest in searching vulnerabilities and how much of this information to share. Authors determine the Perfect Bayesian Equilibrium to extract analytically strategy conditions encouraging information sharing. In \cite{khouzani2014strategic} a firm benefits from losses in another, namely due to exploited bugs. Moreover, they distinguished costs between: direct loss (of compromised firm), common loss by market shrinkage and competitive loss. 

A. K. Eric Luiijf. \cite{Luiijf2015} has analysed the problem of free-riders, i.e. those entities that benefit from the shared information but do not cooperate. To minimize free-riding they propose two approaches: a) provide a quantitative analysis and show the benefits of reciprocity to incentive sharing; b) enforce sharing environments by means of regulations, similar to other works \cite{laube2015mandatory}. In our work, we consider that sharing information may not be always effective, and thus we adopt the first approach, i.e. to study cases in which information sharing benefits the overall network and where it only benefits some of the partners.


One of the main problems when analysing the costs and benefits of information sharing is the experimentation with real data. Whereas most of the proposed works \cite{naghizadeh2016inter}\cite{tosh2015evolutionary}\cite{laube2015mandatory} perform evaluation using analytical methods, Freudiger et al. \cite{freudiger2015controlled} present a controlled data sharing approach and make empirical evaluation using a dataset of suspicious IP addresses. Authors in \cite{freudiger2015controlled} use different similarity metrics to analyse benefits of sharing and compare different sharing strategies: sharing everything or only information about attack entities. Authors in \cite{freudiger2015controlled} rely on a static scenario and provide useful metrics to mathematically predict benefits of info sharing.  By contrast, using a simulated setting we empirically analyse how impacts are propagated through the network, at runtime, to afterwards analyse how information sharing is able to mitigate such impacts in the future.

\section{Background}\label{1sec:background}

This section describes Functional Dependency Network Analysis (FDNA) and game theory, as they are two core disciplines applied in the proposed model. 

\subsection{Functional Dependency Network Analysis}\label{ssec:fdna}

Functional Dependency Network Analysis (FDNA) was proposed first by  Garvey et al. \cite{garvey2009introduction}. They proposed a methodology to assess the impacts derived from loss of supply in one provider to function operability in dependant services. Authors propose two metrics to quantify the dependencies between nodes: the Strength of Dependency and the Criticality of Dependency. Using these metrics, several works have analysed vulnerabilities, impacts and risks in system-of-systems scenarios \cite{oliva2010agent, drabble2012information, guariniello2014communications}.

Few works analyse the benefits of information sharing including dependences among players, as well as the propagation of impacts of cyberattacks. Laube et al.  \cite{laube2015mandatory} refer to dependencies and impact of cyberattacks through direct costs (security breaches happened). By contrast, Hernandez et al. \cite{hernandez2013information} use relationships among nodes inside a sharing community, focusing on knowledge flows and how they increase information value of nodes. We consider that dependencies between partners or entities is a key concept in order to decide whether to share information or not, and thus the proposed model uses functional dependencies. The main goal is to analyse the impact of cyberattacks through services provided in a hypothetical network and how information sharing can contribute to threat mitigation over time.

\subsection{Game theory background}\label{ssec:game-theory}

Game theory relies on four elements to define a game: players, rules or possible actions, information structure and game objective. In a sharing game rules or actions are identified with sharing pure strategies {share, not share}. Games are usually represented in normal form as a pay-off matrix. Pay-offs are numbers representing the outcome, through a measure of quantity or utility that a player gets as a result of playing specific actions \cite{payoffdef}. Pay-offs are the mechanism to reflect the motivations to select pure strategies. Values in a pay-off matrix can be given by constant values or by formulas and are closely related to the goal of the game, thus maximizing or minimizing gained pay-offs.

There are several approaches that develop a game-theory based solution. They go from classical game theory, focused on the analysis of equilibrium among players pay-offs; to evolutionary game theory, focused on the dynamics of strategy changes (in populations). New game-theory approaches are learning game theory \cite{izquierdo2012learning}, where players learn over time based on past decisions of other players, and behavioural game theory \cite{camerer2010behavioral}, based on psychological elements to describe human behaviour. 

Regarding to classification of games, information sharing decisions fit well with the prisoner's dilemma \cite{Luiijf2015}, where cooperative behaviours are not very clear as players have to find a trade-off between benefits and costs of their actions. Moreover, games can be of perfect/imperfect and complete/incomplete information. Perfect information refers to the fact that each player, when making any decision, is informed of all the events that have previously occurred. Complete information refers to the fact that each player has knowledge about the pay-offs and strategies available to the remaining players.

\section{The model}\label{1sec:model}

This section presents the proposed model. Section \ref{ssec:model-overview} presents an overview. Section \ref{ssec:model-desc} describes the model. Sections \ref{ssec:impacts}, \ref{ssec:infoshare} and \ref{ssec:decision-vars} present respectively how cyberattacks propagation is performed, information is shared and decision variables are updated.

\subsection{Model overview}\label{ssec:model-overview}

Organizations and their assets are represented as networks composed of elements called to \textit{nodes}. 
Information/assets owned by nodes have a value. This value represents the information loss (e.g. economical, reputation, resources, etc.) due to the impact of a cyberattack. Accordingly, we consider this value as a combination of Confidentiality, Integrity and Availability principles (CIA) since these properties are at the core of information security. 

\begin{figure}
\centering
\includegraphics[scale=0.4]{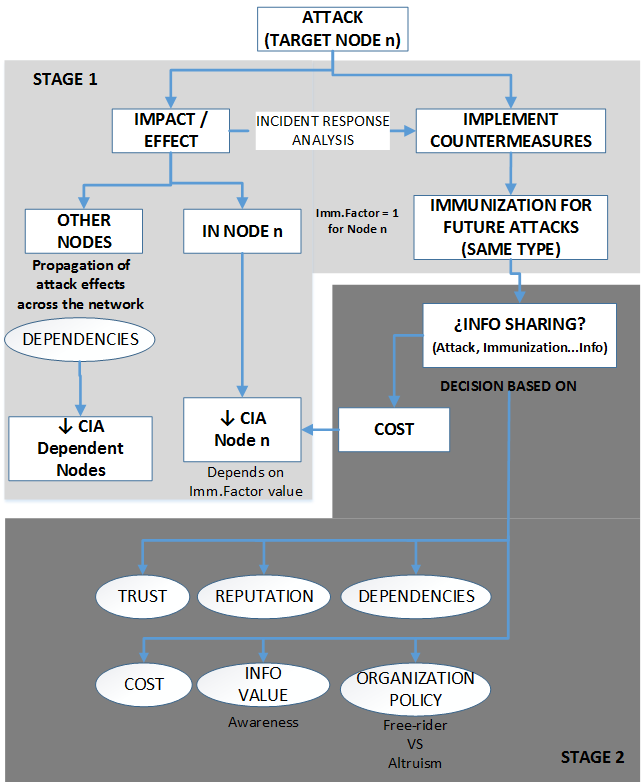}
\caption{Model overview.}
\label{fig:overview}
\end{figure}

The model considers different periods of time (\textit{epochs}) concerning the emergence of a cyberattack. Fig. \ref{fig:overview} presents an overview of the model in which a pair of stages are distinguished: 1) \textit{propagation} of the attack, and 2) information sharing. Stage 1 shows that when a node is targeted by an attack, its CIA value decrease proportionally to the impact of the cyberattack. As a consequence of an attack, dependant nodes on the targeted node are also affected according to service levels agreed among nodes. This process is what we call \textit{propagation of cyberattacks}.

Upon receiving an attack, the targeted node is able to develop countermeasures (e.g. due to an incident investigation), thus being \textit{immunized} for future attacks. Then, attacked nodes must decide whether to share the information about attacks (and the associated countermeasures) with other nodes. 

Information sharing steps are represented in the Stage 2 of Fig. \ref{fig:overview}. The sharing decision is based on several variables and environmental conditions. On the one hand, if one node  decides to share information, it will incur a cost which is related to the risk of unwanted disclosure of shared information and privacy loss. Thus, this cost directly affects CIA value of the node with which information is shared. On the other hand, sharing may provide two main benefits: (1) to raise cybersecurity awareness level through immunization factor, (2) to improve enterprise reputation. In general, sharing decisions are based on several criteria and have many associated variables: with whom to share, cost of sharing, dependencies, trust and reputation among nodes, organization policy, information properties, level of resilience and knowledge acquired in sharing processes, among others. The problem is formulated as a trade-off between costs and benefits of sharing information. As we show in Section \ref{1sec:experimentation}, simulations with this model can be used to analyse the scenarios where different sharing strategies are better, regarding different conditions and variables like the presence of free-riders.

\subsection{Model description}\label{ssec:model-desc}
Let $\Omega = (A,V,C,Y,I,S,T,R,W)$ be the set of variables representing the state of the model at each epoch. Sharing decisions and results obtained at each epoch are based on the values of variables of $\Omega$. In the following we describe these variables.

\textbf{Services and dependencies of the network ($A$)}

The model considers a network composed by nodes representing some ICT infrastructure. Each node encapsulates a relevant element or service having functional dependencies with others (elements or services) inside/ outside the organization networks. A link between two nodes represents the dependency level between those nodes. The network can be viewed as a labelled graph, represented as an adjacency matrix $A$ of size $n \times n$, being $n$ the number of nodes . An element $a_{i,j}$ in matrix $A$ represents the level of service that node $i$ offers to node $j$, or the degree of dependency of node $j$ on node $i$. Values of elements range from 0 to 1, where 0 means no service/dependency and 1 means as total dependency for one node to other. 

\textbf{CIA value of the nodes ($V$)}
Each node has a CIA value which bases on the three traditional information security properties: confidentiality (C), integrity (I) and availability (A). $V$ is a vector of size $n$ where each element $v_i$ is the value of the node $i$ inside the network. The values range from 0 to 1, where 0 means that the node has no value (i.e. useless asset) and 1 means that the node is an important asset for an entity. While this is an interesting and complex area on risk analysis and asset management, establishing asset values is out of the scope of this work.

\textbf{Costs of information sharing ($C$)}

When an entity or organization shares information, it may incur a cost. 
Costs of sharing are described by a vector $C$ of size $n$, where each element $c_i$ is the cost of sharing for node $i$. Similarly to giving values to assets, quantifying costs is not straightforward and it depends on the scenario of application. The cost of information sharing has a direct effect on the CIA value of nodes (e.g. due to loss of confidentiality or privacy). To simplify the model, in our current prototype these costs are calculated using a parameter $k \in [0,1]$ that represents the percentage of loss in the CIA value of the node due to sharing information. Thus, the cost of sharing is directly proportional to the CIA value of the nodes. Summarizing, values in the vector $C$  are calculated as $c_i = k \cdot v_i$, where $v_i$ is the CIA value of node $i$ and $c_i$ its cost of sharing.

\textbf{Set of possible cyberattacks ($Y$)}

A cyberattack is any attempt to compromise the confidentiality, integrity or availability of an asset (node) in an organization. The caused damage can be viewed as an economic impact, but in our case, the impact decreases CIA value of the targeted node. Each cyberattack has specific properties and the organization should implement the correct countermeasures to solve problems as soon as possible. The set of cyberattacks is described by a vector $Y$ of size $n$,  where each element $Y_i$ is the cyberattack received by node $i$ in a given epoch. 
The model manages $m$ possible cyberattaks\footnote{We set a limited number of cyberattacks in our current prototype implementation, but these cyberattack may be viewed as a zero day if none of the nodes are immunized against them}, and it is assumed that cyberattacks have a default impact associated, that we note as $D$. The values of $m$ and $D$ can be modified during experimentation to simulate different attacks scenarios.

Attacks propagation causes a pair of different damages. First, the targeted node suffers a direct impact which can be mitigated if the node was prepared for it (\textit{immunized}), thus it depends on the immunization factor of this node for this attack, as it is explained below. Second, a \textit{related} impact is applied on nodes that directly or indirectly depend on the targeted node. In this case, the CIA value is decreased according to the degree of dependency between the nodes and the direct impact of the targeted node.

\textbf{Immunization factors ($I$)}

An immunization factor indicates how well a node is prepared against a cyberattack, and thus, how it can mitigate its impact. It represents any mechanism or countermeasure such as knowledge about this attack (IDS rules, IoCs, blacklisting IPs, etc.), a piece of software (antivirus, SIEM), organization policies, etc. The immunization factor is specific for each attack and each node, and it is represented by a matrix $I$ of size $m \times n$, where $m$ is the number of available cyberattacks and $n$ is the number of nodes. An element $I_{p,i} \in [0,1]$ represents the degree (factor) by which node $i$ is immunized for cyberattack $p$. Accordingly, a value $I_{p,i}=0$  means that node $i$ is not immunized for cyberattack $p$, while a value $I_{p,i}=1$ indicates that node $i$ is totally immunized for cyberattack $p$ and it will have no impact on CIA value of $i$. 

An important point is the difference between the \textit{direct} and \textit{related} attack impact, as well as the relationship with their immunization factor. Once a node receives a cyberattack, it impacts its CIA value, only mitigated according to the immunization factor of such node. However, if this node is not immunized against the attack, the corresponding impact will be propagated across the network based on dependences, no matter the immunization of other nodes against this cyberattack. As a result, sharing information becomes essential if a node does not want to be affected by cyberattacks suffered by other nodes.

\textbf{Sharing policy ($S$)}

Sharing policies determine nodes behaviour to decide with whom to share or not to share. The decision bases on the properties of each node, network conditions and variables in $\Omega$. Each node chooses a fixed sharing policy along the whole game to select a pure strategy, namely, share or not share. These policies are described by $S$. $S$ is a vector where $S_i$ is the sharing policy of the node $i$. Examples of sharing policies are "\textit{Share only with those on which I directly depend}" or "\textit{Share with those nodes that provide more services}".


\textbf{Trust among nodes ($T$)}

In the cooperative cyber defense scenario, trust is a key aspect, since it is used to define collaborative security models \cite{meng2015collaborative}. Trust represents how nodes trust each other \cite{gambetta2000can} and it is described by a matrix $T$ of size $n \times n$. An element $T_{i,j} \in [0,1]$ is the trust value that node $i$ has on node $j$.  $T_{i,j}=0$ means node $i$ has no trust in node $j$ and $T_{i,j}=1$ means node $i$ has full trust in node $j$. Trust between nodes increments or decrements the cost of sharing, i.e. sharing with those nodes that are more trustful is less costly.

\textbf{Reputation among nodes ($R$)}

Trust and reputation are related but different concepts \cite{gambetta2000can}. On the one hand, trust is the subjective probability that an agent will carry out a specific task as expected. On the other hand, reputation represents the set of past opinions received by others users, that is, expectancy of behaviour based on past interactions \cite{abdul2000supporting}. Our approach represent reputation through a matrix $R$ of size $n \times n$ where an element $R_{i,j} \in [0,1]$ is the reputation value that node $i$ obtain for node $j$.  $R_{i,j}=0$ means node $i$ receive no reputation from node $j$ and $R_{i,j}=1$ means node $i$ receive full reputation from node $j$.

\textbf{Awareness ($W$)}

Awareness represents the degree of useful information received by a particular node, i.e. information which was unknown by the receiver and thus it increases the general awareness of the node at each epoch. We represent awareness through a matrix $W$ of size $n \times n$ where an element $W_{i,j} \in {0,1}$ is the awareness that node $i$ obtains for node $j$.  $W_{i,j}=0$ means node $i$ receives no valuable information from node $j$ and $R_{i,j}=1$ means node $i$ receives valuable information from node $j$. In this first approach we closely relate the awareness with immunization factors, since node $i$ gets $W_{i,j} = 1$ from node $j$ only if $i$ gets new immunization factors for some cyberattack $p$ not yet immunized, thus $I_{p,i}=1$ for cyberattack $p$ received by node $j$.\\



Algorithm \ref{alg:simmodel} shows the sequence of actions executed in the model. As input, it receives an initial state of the system represented by $\Omega$ as explained above, and as output, it provides a set of metrics and features regarding the final state and the sequence of actions from the simulation (e.g. which nodes have shared information, which nodes have been targeted by cyberattacks, etc.). These metrics are used to perform an analysis both in the network and in particular nodes. \\ 

It is important to note in line 3 of Algorithm \ref{alg:simmodel} we generate the attack vector $Y$, establishing the nodes that are targeted at each epoch, and which attacks inside the set of attacks are used to impact the nodes according to default attack impacts in $D$. 



\begin{algorithm}
\caption{Cyber-attack Propagation and Information Sharing Simulation Model}\label{alg:simmodel}
\begin{algorithmic}[1]
\Procedure{CyberModel}{$\Omega = (A,V,C,Y,I,S,T,R,W)$}
\For {$t:=1 \to MAX\_EPOCH$}
\State $Y \gets$ Set Attack Vector
\State Calculate impact in nodes according to $Y$ and $D$
\State $I \gets 1$ Set Immunization Factor = 1 
\State Propagate impacts through the network
\State Set Sharing strategies
\State Play Sharing Game 
\State Update CIA values according to sharing policies
\State Update Reputation regarding sharing decisions
\EndFor
\EndProcedure
\end{algorithmic}
\end{algorithm}

\subsection{Propagating cyber-attacks impacts}\label{ssec:impacts}
In the first step of the model, once a node is under attack, dependant nodes are somehow affected. The propagation of these impacts is the first process computed in each epoch (from line 4 in Algorithm\ref{alg:simmodel}). 

On one hand, the direct impact on targeted node decreases the value of this node according to Eq. \ref{eq:impact}, 

\begin{equation}\label{eq:impact}\forall i \in V, V_i^t = V_i^{t-1} - \parallel V_i^{t-1} \cdot (D_{Y_i^t} \cdot (1 - I_{Y_i^t,i})) \parallel \end{equation}

Here, $Y_i^t$ is the cyberattack received by node $i$ in epoch $t$, and $D_{Y_i^t}$ is the default impact of such an attack according to initialized values as described in previous section. $I_{Y_i^t,i}$ is the immunization factor of node $i$ for the cyberattack $Y_i^t$. Note that if $I_{Y_i^t,i}$ is 0, then node $i$ is fully impacted by all $D_{Y_i^t}$, and if $I_{Y_i^t,i}$ is 1, then node $i$ is not impacted at all.

Without loss of generality, in the current implementation of the model nodes that are directly attacked get an immunization factor of 1 for that specific attack (thus $I_{Y_i^t,i} = 1$). This represents a total immunization for future attacks of the same type, due to a perfect incident response procedure (we discuss this assumption in Section \ref{ssec:assumptions}).

On the other hand, the impact of the cyberattack is propagated across the network (line 6 in algorithm \ref{alg:simmodel}). Propagation is carried out on both direct and indirect dependently nodes. Direct dependent nodes are well represented in dependency network matrix $A$, while indirect dependent nodes are more difficult to get. To address this issue, we previously calculate indirect services matrix $B$, also using a Deep First Search algorithm on $A$. With this indirect service matrix we can calculate impact of cyberattacks and the new CIA value for each node in the network, as showed in Eq. \ref{eq:propag}:

\begin{equation}
\label{eq:propag}
\forall j \in V, V_j^t = V_j^{t-1} - \parallel V_j^{t-1} \cdot (D_{Y_i^t} \cdot (1 - I_{Y_i^t,i})) \cdot (B_{i,j}) \parallel 
\end{equation}

Here, $i$ is the node directly attacked, and $j$ is the dependant node impacted. $B_{i,j}$ is the service weight that node $i$ offers direct or indirectly to node $j$. As described before, immunization factor does not reduce the impact of the propagated effect of one attack, that is, if one node $j$ receives an impact derived from an attack on other node $i$, the effect on node $j$ only depends on the effect on node $i$ and the service weight defined in $B_{i,j}$.

\subsection{Information sharing game}\label{ssec:infoshare}
The second step of the model is to decide if nodes under attack share information or not. In this step, identified by lines 7 and 8 of in algorithm \ref{alg:simmodel}, sharing strategies are established according to sharing policies and apply them to calculate pay-offs (benefits and costs) at each epoch. 

\subsubsection{Game definition}\label{ssec:gamedef}
Players are the nodes of the network and the set of pure strategies is \{Share,Not Share\}. According to the number of players, our approach proposes a multiple two-players (pairs) game. Thus, we define an iterative 2-nodes multiple games, with  $n \cdot (n-1)$ different games played with different results at each epoch. Moreover, the proposed game is dynamic due to the fact that participants play stages of the same game over time. So we propose an instance of the Iterated Prisoner's Dilemma as decisions can vary over time and players have to decide their strategies without communicating them to others under free-riding conditions.

Non-cooperative and inefficient games can lead to efficient trade-off through repetition \cite{naghizadeh2016inter}, but it is paramount to know if games are under perfect or imperfect information, and complete or incomplete information conditions, as well. Our approach follows an imperfect and complete information game, since every node knows about pay-offs configuration and strategies available to other players, but they do not know every action performed by the others (imperfect information).

Additionally, shown in Table \ref{table:pay-offs}, this is a non-zero sum game, while pay-off gained by one node is not exactly what the other losses. Furthermore, the game is asymmetric since two players can get different pay-offs even when applying the same strategy. This is because pay-offs depend on particular trust, information properties and the degree of dependency among nodes. Even though the prisoner's dilemma take symmetric configuration, it can take an asymmetric form as presented in \cite{wang2013asymmetric}, and it is important to note that some works \cite{beckenkamp2007cooperation} point out that asymmetry reduces cooperation rates in prisoner's dilemma games.


\subsubsection{Sharing strategy}\label{ssec:sharing-stg}

Selecting the sharing strategy for each node is one of the most important decisions to take in cooperative scenarios.
We distinguish between sharing strategy and sharing policy concepts. Sharing strategies refer to share or not to share, that is the pure strategies. On the other hand, sharing policies correspond to \textit{"strategies"} that players use to decide what pure strategy to play. Sharing policies can be supported from basic to very complex decision processes. Classical and most common sharing policies applied to Iterative Prisoner's dilemma for behavioural analysis of players are \cite{li2007design}: AIIC (Always Cooperate); AIID (Defects on every move); RAND (Random player); TIT (Tit for Tat), that is, cooperate on the first move, then copies the opponent's last move; and Grim (Grim Trigger), cooperates, until the opponent defects, thereafter always defect. Sharing policies are particular for each scenario, as we will see in Section \ref{1sec:experimentation}.



\subsubsection{Calculating pay-offs}\label{ssec:sharing-payoffs}
Pay-off functions depend on reputation, awareness, trust and cost of sharing sensible information with others. Pay-off matrix is shown in Table \ref{table:pay-offs}, where $U_A$ and $U_B$ are respectively the pay-offs (utilities) of $A$ and $B$ obtained for each of four possible combination of players' actions. There are mainly two factors in each equation, first, values allocating benefits, and secondly, values allocating costs according to the chosen strategy. In general terms, it indicates what players win and what players loss. Regarding the notation, $R_{a\leftarrow B}$ is the reputation that node $A$ receives from $B$, $T_{A\rightarrow B}$ represents trust from node $A$ to node $B$ and $C_A$ is the cost of sharing for node $A$. These values are extracted from variables defined in Section \ref{ssec:model-desc}. 

$W_{A\leftarrow B}$ represents the awareness degree gained by node $A$ from node $B$. 
It is calculated as $W_{A\leftarrow B} = 1 \leftrightarrow $ ($B$ shares with $A$) $\wedge$ ($B$ is under attack) $\wedge$ ($A$ is not immunized for the attack received by $B$). That is, node $A$ gets immunized to a given attack which it was not prepared for, due to information received from node $B$. 

Also note how trust directly affects the cost of sharing. $\frac{C_A}{T_{A\rightarrow B}}$ means that the higher the trust from $A$ to $B$, the lower cost of sharing from $B$ to $A$ and vice versa.

\begin{table*}[ht]
\small
\centering
\caption{Pay-off matrix for information sharing game}
\label{table:pay-offs}
\begin{tabular}{|p{1cm}|p{2cm}|p{5cm}|p{5cm}|}
\hline
                   &           & \multicolumn{2}{c|}{B}                                                                                                                                        \\ \hline
                   &           & Share                                                                         & Not Share                                                                     \\ \hline
\multirow{2}{*}{A} 
& Share     
& $U_A: (R_{A\leftarrow B} + W_{A\leftarrow B}) - (\frac{C_A}{T_{A\rightarrow B}})$ 
\newline \newline 
$U_B: (R_{B\leftarrow A} + W_{B\leftarrow A}) - (\frac{C_B}{T_{B\rightarrow A}})$ 
\newline 
& $U_A: R_{A\leftarrow B} - (\frac{C_A}{T_{A\rightarrow B}})$
\newline \newline 
$U_B: (C_B + W_{B\leftarrow A}) - R_{B\leftarrow A}$
\\ \cline{2-4} 
& Not Share 
& $U_A: (C_A + W_{A\leftarrow B}) - R_{A\leftarrow B}$,                           
\newline \newline 
$U_B: R_{B\leftarrow A} - (\frac{C_B}{T_{B\rightarrow A}})$
& $U_A: C_A - (R_{A\leftarrow B} + W_{A\leftarrow B})$,                 
\newline \newline 
$U_B: C_B - (R_{B\leftarrow A} + W_{B\leftarrow A})$ 
           
\\ \hline
\end{tabular}
\end{table*}

\subsection{Updating decision variables}\label{ssec:decision-vars}

Once attacks have been propagated and the information has been (or not) shared, the final step  is to update CIA value and reputation of all nodes.

In addition to the decrease of CIA value due to attack impacts, sharing information has an effect on the CIA value as well, that is, the cost of information sharing as shown in Eq. \ref{eq:cost-sh}. This reduction due to information sharing is only applied when a node has been targeted by a cyberattack, it is not previously immunized, it has shared information with any other node and the mean pay-off obtained by sharing information is less than 0.

\begin{equation}
\label{eq:cost-sh}
\forall i \epsilon V, V_i^t = V_i^{t-1} - C_i
\end{equation}

Also, reputation obtained for each node is updated at each epoch. As algorithm \ref{alg:repupdate} shows, the process for updating reputation depends on previous reputation scores,  sharing actions, awareness obtained and whether nodes are under attack or not. In order to prevent free-riding behaviours, we identify three cases, formalized in lines 4 to 11 in Algorithm \ref{alg:repupdate}: (1) if node $j$ increase its awareness by information shared by node $i$, that is, $j$ gets immunized by $i$, then $j$ rewards the $i$ reputation score by a constant $K_{reward}$; (2) if node $j$ does not increase its own awareness by $i$ but $i$ is determined to share, then it is difficult to know if the node $i$ is a free-rider or the information provided is not useful. In this case, we propose to do not modify previous reputation value; (3)  if node $j$ does not increase its own awareness by $i$ and $i$ is not determined to share (free-riding behaviour), then reputation score of $i$ decreases by punishing with $K_{punish}$. 

\begin{algorithm}
\small
\caption{Updating reputation scores procedure}\label{alg:repupdate}
\begin{algorithmic}[1]
\For {$n:=1 \to n$}
	\For {$j:=1 \to n$}
		\If {$j\neq n$}
			\If {$(Awareness(j,n) > 0)$}\\
				$R_{n\leftarrow j}^t =  R_{n\leftarrow j}^{t-1} + (K_{reward} \cdot R_{n\leftarrow j}^{t-1})$
			\Else
				\If {SharingStrategy(n,j)=Share} \\
				$R_{n\leftarrow j}^t =  R_{n\leftarrow j}^{t-1}$						
				\Else\\
				$R_{n\leftarrow j}^t =  R_{n\leftarrow j}^{t-1} - (K_{punish} \cdot R_{n\leftarrow j}^{t-1})$											
				\EndIf
			\EndIf			
		\EndIf
	\EndFor
\EndFor

\end{algorithmic}
\end{algorithm}


\subsection{Assumptions}\label{ssec:assumptions}
Due to limitations on simulation, during the definition of our model, we have made some assumptions that may be subject of discussion in real settings. First, value of nodes always decrease over time due to the impact of cyberattacks and the costs of information sharing. However, we do not consider mechanisms that may increase this value, such as contingency plans or asset restoration. Secondly, when a node has been targeted by an attack, we assume that it performs a proper incident response investigation and thus it implements countermeasures, gaining an immunization factor of 1. In this way, from this moment on the node will be immunized to such cyberattack. 


\section{Experimentation}\label{1sec:experimentation}

This section presents a simulation framework based on the model described in Section \ref{1sec:model}, whose prototype have been implemented in Octave\cite{octave}. Using this prototype, a few case studies are analysed by means of the metrics provided by the simulation. First, goals and general conditions of the simulation framework are introduced in Section \ref{ssec:simulation-frame}. Second, Section \ref{ssec:experiment-setup} presents the analysed scenarios and specific settings. In Section \ref{ssec:metrics} we describe used metrics and present analysis results for both general welfare and specific nodes viewpoints.

\begin{figure*}
\centering
\includegraphics[scale=0.6]{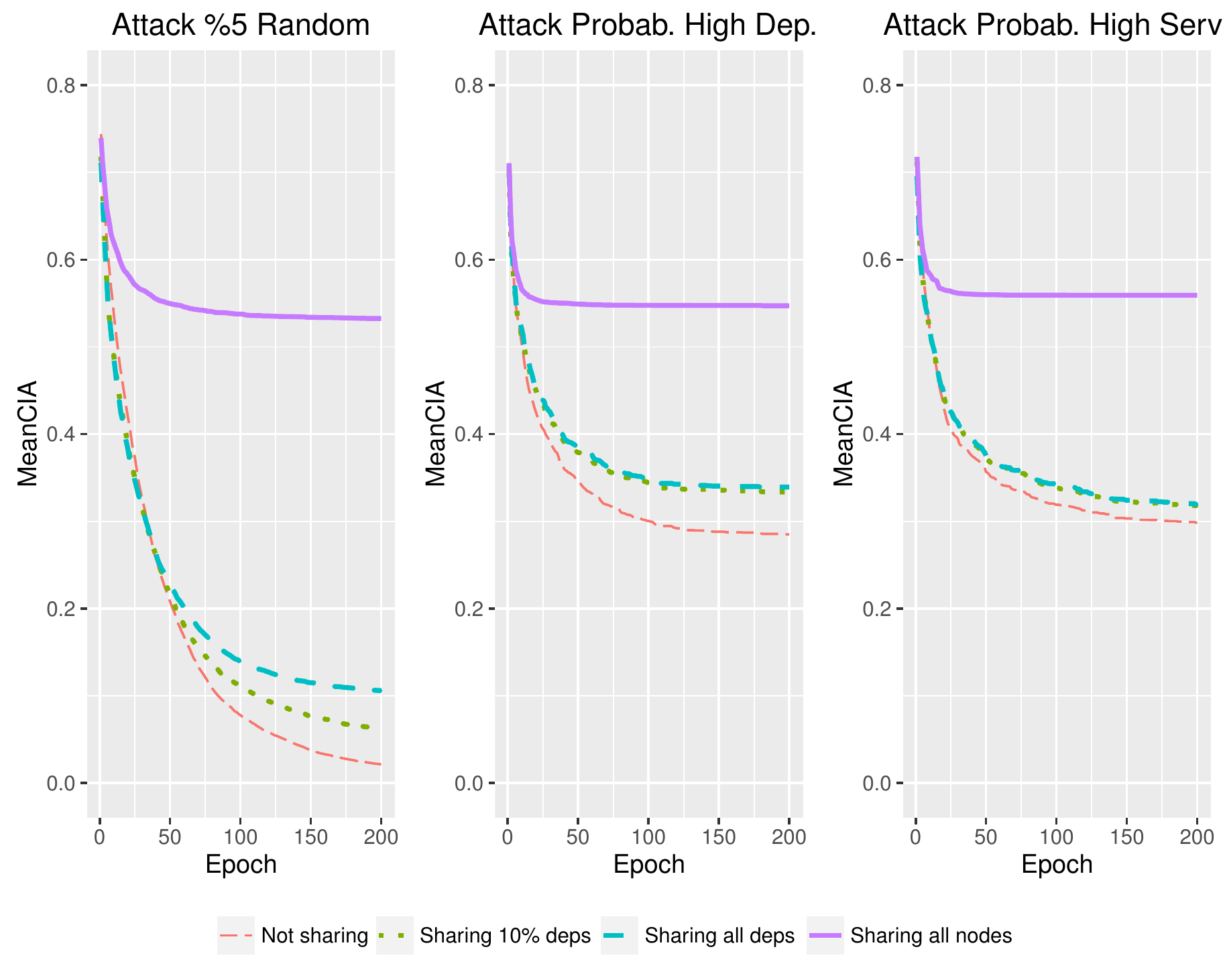}
\caption{Evolution of mean CIA value of nodes for each attack and sharing scenario}
\label{fm}
\end{figure*}

\subsection{Simulation framework}\label{ssec:simulation-frame}

Empirical evaluation in real large-scale cybersecurity information sharing environments is hard to carry out. This is due to the highly diverse and dynamic conditions of the scenarios \cite{Spijkervet2014}. Thus, we have implemented a simulation framework using GNU Octave \cite{octave} based on the proposed model. This simulation framework allows to simulate specific cyberattacks and information sharing scenarios, as well as the study of the behaviour and evolution of nodes (assets) over time. The main goal is to show how the model can help in decision-making problems by analysing test cases.

We aim to build a flexible framework with configurable settings, namely network size and topology, attack scenarios, sharing policies and initial trust, reputation and CIA.



To foster research on information sharing, we provide a open-source version of the prototype in our github repository\footnote{https://github.com/rguseg/infosh-framework}

\subsection{Experimental set-up}\label{ssec:experiment-setup}
We carry out a series of experiments to analyse specific network topology behaviours over time based on Monte Carlo simulations method. We aim to compare network and nodes evolution according to different sharing policies in three different adversarial models. Hereafter we describe our experimental set-up. It is important to note that the main goal of our experimental work is to show up the benefits of using the proposed model and how it can be used in different case studies. Established settings may not represent real scenarios, but they aim to simulate general idiosyncrasies of current networks.

\textbf{Network sampling}. We set a random scale-free directed network composed by 50 nodes. We consider that 50 nodes represent a medium-sized sharing community. Also, we choose a scale-free network since it has a similar topology than the Internet \cite{boccaletti2006complex}. Note that in scale-free networks some nodes are highly connected, while most of the nodes have low connectivity. In the proposed network one node has an input degree of 14, and most of them have 2 or 3. Output degrees have similar properties. Regarding weight of dependencies, we simulate a network where dependencies values are fixed to 0.5 among every connected node, so each node is equally dependent of its neighbours.


\textbf{Attack scenario} 
It presents the degree of threat to which a sharing community is exposed, and it is determined by the different cyberattacks, their impact, their frequency and which nodes are targeted. We randomly chose attacks from a predefined catalogue with an associated impact. 

In particular, we assume the existence of a catalogue composed of 10 attacks $(m=10)$ to give the simulation enough variability. The impact of each cyberattack follows a normal distribution with mean $0.4$ and standard deviation of $0.2$. It means that each cyberattacks has an impact between $0.2$ and $0.6$ on the targeted node. 

At each epoch, 30\% of the attacks from the catalogue are randomly selected. Then, a subset (vector) of nodes $Y$ become the targets. Selection of targeted nodes depend on many specific conditions, and it may be variable in real settings (e.g. the knowledge of the adversary about the network topology or the available vulnerabilities to exploit). In the experimentation we use the following criteria: 

\begin{itemize}
 
\item Random selection. We randomly select 5\% of nodes of the network to be attacked.
\item Attack to those with higher dependencies. We first calculate the number of inputs degrees for each node and sort them in descendent order. For each node, we estimate the probability of being attacked as $p = \frac{Degree_{Input}}{n}$, where $n$ is the number of nodes in the network. Consequently, the higher the weight of inputs (dependencies), the higher the likelihood of being targeted. 
\item Attack to those providing highest number of services. It applies the same procedure as for dependencies, but taking nodes outputs degrees instead of inputs degrees. The probability of being attacked is estimated as $p = \frac{Degree_{Output}}{n}$. 
\end{itemize}

When the degree of a node is 0, it means that it has no dependencies or services to offer. Still yet, they may be target of cyberattacks, so in these cases we set a default (minimum) probability of being targeted equal to 2 \% ($p=0.02$).

\textbf{Sharing policies}. 
We set-up four available sharing policies or each simulation as presented in Section \ref{ssec:infoshare}. Since our goal is to show how the model helps in deciding which sharing strategies are better (independently of local policies), in our experimental work sharing policies are static and global, i.e. they do not change over time and are the same for all the nodes. These strategies are: a) no-one shares; b) nodes share with 10\% of nodes who they most depends on (weight of dependencies); c) nodes share with all the nodes they depends on (100\%); d) nodes share with all nodes in the network (broadcast).\\

\textbf{Initialization of values}.
CIA value of nodes is set to 0.8 in a range [0,1]. Setting CIA values to 1 would mean that a node is the most valuable asset for an organization, but we do not aim to simulate most critical assets in our experiments.
Given the difficulties of establishing trust values \cite{granatyr2015trust}, initial trust values are set to 0.5, being in the range [0,1]. This decision represents a neutral position which remains static over time.\\


\textbf{Other Parameters}. 
For each particular scenario, we run 30 simulations of 200 epochs. The percentage of loss in CIA value $k$ of each node due to information sharing (i.e. the cost) is set to 0.2 and the level of punishment ($K_{punish}$) and reward ($K_{reward}$) are set to 0.3 respectively.

\subsection{Results}\label{ssec:metrics}

We focused the analysis in two areas. First, we focus on the evolution of the general welfare of the network during time, due to the application of different sharing policies. Second, we aim at analysing and identifying particular nodes that have different benefits under similar conditions, so as to extract useful information from them. To this end, we define three metrics, named \textbf{MeanCIA}, \textbf{Gain} and \textbf{Information Quality}, and then we use these metrics to carry out the analyses.

\begin{theorem}
$\bar{V^{t}}$ is the \textbf{MeanCIA} metric that represents the general welfare of the network in epoch $t$. It is calculated as the average value of CIA of all nodes and all simulations. 
\end{theorem}

The \textbf{MeanCIA} is calculated as the average value of CIA of all nodes and simulations, as showed in Eq. \ref{eq:meancia}.


\begin{equation}\label{eq:meancia}\bar{V^{t}} = \frac{1}{sims} \cdot \mathop{\sum}_{s=1}^{sims} (\frac{1}{n} \cdot \mathop{\sum}_{i=1}^{n} V_{i}^{s,t}) \end{equation}
where $t$ is the epoch which we want to process, $sims$ is the number of running simulations, $n$ is the number of network nodes, then $V_{i}^{s,t}$ is the CIA value of node $i$ in simulation $s$ and epoch $t$.

\begin{theorem}
$G_i^t$ is the \textbf{Gain} metric that indicates the degree of CIA value gained or lost by each node because of information sharing. It is calculated  as the difference in the CIA values of each node after applying two policies (sharing with all the nodes and not sharing) in two similar scenarios with the same conditions.
\end{theorem}
The \textbf{Gain} is calculated as showed in Eq. \ref{eq:gain}. 
\begin{equation}
\label{eq:gain}G_i^t = V_i^{sh2,t} - V_i^{sh1,t}
\end{equation}

where $t$ is the epoch which we want to process, $i$ is the node, $V_i^{sh1}$ is the CIA value of node $i$ while applying the sharing policy $sh1$ (not sharing), and $V_i^{sh2}$ is the CIA value of node $i$ while applying the sharing policy $sh2$ (sharing with all the nodes).

\begin{theorem}
$Q$ is the \textbf{Information Quality} metric that represents the amount of information that is useful to increase the awareness of the nodes. Concretely, $Q_i^{in}$ is the amount of quality information sent and $Q_i^{out}$ is the amount of quality of information received by node $i$ .
\end{theorem}

Based on these metrics, we next present the results obtained and the analysis performed on the three attack scenarios and applying the different sharing policies.

\subsubsection{Analysis of the general welfare of the network}\label{ssec:analysis-net}
In this section, we provide an analysis of welfare evolution over time for
the four sharing policies applied in each of the three attack scenarios as described in Section \ref{ssec:experiment-setup}, thus a total of 12 different scenarios.

Fig. \ref{fm} shows time evolution of \textbf{MeanCIA} during 200 epochs for the four sharing policies. 
It can be observed that at the beginning of the simulation, all sharing policies have rather similar \textbf{MeanCIA}, but it decreases faster when applying policies that do not share or perform selective sharing. Besides, applying sharing policies based on dependences do not produce too much benefits. Concretely, sharing with 10\% of dependent nodes only improves not sharing policies in a ratio of 5\%, 6\% and 2\% on the three attack scenarios respectively, while sharing with all the dependent nodes only improves 10\%, 7\% and 3\%. Sharing with all nodes in the networks gives improvements rates over not sharing of 64\%, 33\% and 32\%. Thus, it can be concluded that only the policy of sharing with all nodes is substantially beneficial for the general welfare of the network in the terms and conditions established during our experimentation.

Regarding the effects of different attack scenarios on the \textbf{MeanCIA}, we can observe that, in general, ``Attack 5\% Random" has higher impacts than ``Attack Probab High Deps" and ``Attack Probab High Servs.", no matter what the information sharing policy is being applied. Thus, from the adversarial point of view, it is better to attack randomly if she knows that no information sharing or sharing only with dependent nodes, is being applied.


\subsubsection{Analysis and identification of critical nodes}\label{ssec:analysis-nodes}

In this section, we provide an analysis to detect nodes that are more important to the community in terms of the \textbf{Information Quality} as presented above. Thus, the amount of information shared to others that actually increase their awareness, i.e. this information was not known previously in the community. We analyse these relevant nodes according to the \textbf{Gain} metric defined above.

First, we calculate $G_i^t$ for every node $i$ in the network and simulation, focusing only in the last epoch $t=200$. We only compare sharing scenarios with more relevant differences found in Fig. \ref{fm}, that is, scenarios applying sharing policies a) and d) as defined en Section \ref{ssec:experiment-setup}. This means \emph{sh1} and \textit{sh2} described in Eq. \ref{eq:gain} are \textit{not sharing }and \textit{sharing with all nodes}, respectively. Nodes that benefit from scenario \textit{sh2} respect to scenario \emph{sh1} are those with $G_i^{200} > 0$. Nodes that do not benefit from scenario \textit{sh2} respect to scenario \emph{sh1} are those with $G_i^{200} \leq 0$.

\begin{figure}
\centering
\includegraphics[width=\columnwidth]{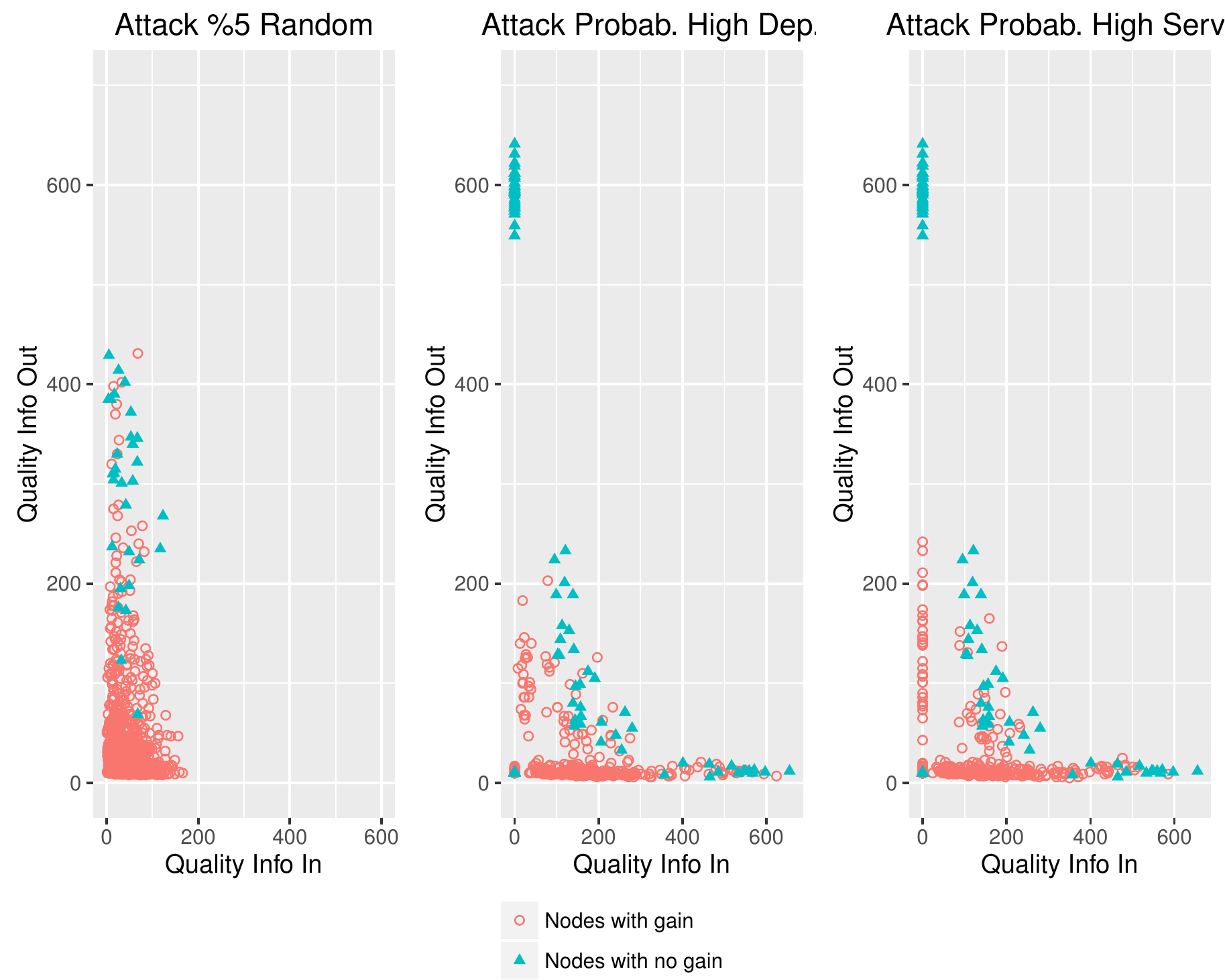}
\caption{Distribution of nodes that gain (red circles) and lose out (blue triangles) because of information sharing, in terms of number of pieces of quality information shared and received}
\label{fig:figure-awareness}
\end{figure}

Second, we conduct the analysis in terms on how many pieces of \textit{quality} information are sent $Q_i^{out}$ and received $Q_i^{in}$ by each node. Fig. \ref{fig:figure-awareness} shows the distribution of nodes in terms of how many pieces of \textit{quality} information is provided (y-axis = $Q_i^{out}$) and received (x-axis = $Q_i^{in}$). Red triangles represent those nodes that do not obtain benefits because of information sharing ($G_i^{200} \leq 0$), and blue circles represent the opposite ($G_i^{200} > 0$). In general, it can be observed that nodes that do not obtain any benefit usually provide more pieces of \textit{quality} information than they receive. This may occur because they receive attacks that are new to the network (e.g. zero day exploits) and afterwards they notify (and immunize) the remainder nodes about them, but then they do not receive information regarding other attacks. While this obviously depends on the specific settings and the attack scenario, the analysis provided with this simulation framework allows the identification of nodes whose information is relevant for the health of the entire network. Moreover, since these nodes may not benefit from the sharing community, in real settings they could be rewarded by other means (e.g. providing extra economical benefits) to incentive their cooperativeness.

\section{Conclusions}\label{1sec:conclusions}
Attack prevention and detection are essential task in cybersecurity management. Organizations suffer multiple cyberattacks and information sharing can help to develop early prevention mechanisms. However, organizations are not willing to share information unless incentives are achieved. In this regard, this paper presents a model for cybersecurity information sharing among dependently organizations being impacted for different cyberattacks. Functional Dependency Network Analysis is used for attacks propagation and game theory for information sharing management. A framework has been developed and the model has been tested in a particular scenarios. Results using the simulation framework suggest that info sharing generally improves the general welfare. Moreover, we elucidate that nodes that receive new attacks in the network provide information with higher quality even though they do not benefit from the sharing community, so they should be rewarded and motivated to share such information. In general, our experimental work shows that the proposed model can help to simulate network conditions and adversarial settings to analyse beneficial sharing policies, both in terms of particular nodes and in terms of the general welfare of the sharing community. It can be used as a part of a decision support system or during countermeasure allocation processes.

\section{Acknowledgements}\label{1sec:acknowledges}
This work was partially supported by the MINECO grant TIN2013-46469-R and the CAM grant S2013/ICE-3095 CIBERDINE -CM funded by Madrid Autonomous Community and co-funded by European funds.

%
\bibliographystyle{abbrv}
\bibliography{infosh}  

\begin{thebibliography}{10}

\bibitem{abdul2000supporting}
A.~Abdul-Rahman and S.~Hailes.
\newblock Supporting trust in virtual communities.
\newblock In {\em System Sciences, 2000. Proceedings of the 33rd Annual Hawaii
  International Conference on}, pages 9--pp. IEEE, 2000.

\bibitem{beckenkamp2007cooperation}
M.~Beckenkamp, H.~Hennig-Schmidt, and F.~P. Maier-Rigaud.
\newblock Cooperation in symmetric and asymmetric prisoner's dilemma games.
\newblock {\em MPI Collective Goods Preprint}, (2006/25), 2007.

\bibitem{boccaletti2006complex}
S.~Boccaletti, V.~Latora, Y.~Moreno, M.~Chavez, and D.-U. Hwang.
\newblock Complex networks: Structure and dynamics.
\newblock {\em Physics reports}, 424(4):175--308, 2006.

\bibitem{camerer2010behavioral}
C.~Camerer.
\newblock {\em Behavioral game theory}.
\newblock New Age International, 2010.

\bibitem{drabble2012information}
B.~Drabble.
\newblock Information propagation through a dependency network model.
\newblock In {\em Collaboration Technologies and Systems (CTS), 2012
  International Conference on}, pages 266--272. IEEE, 2012.

\bibitem{octave}
J.~W. Eaton, D.~Bateman, S.~Hauberg, and R.~Wehbring.
\newblock {\em {GNU Octave} version 4.0.0 manual: a high-level interactive
  language for numerical computations}.
\newblock 2015.

\bibitem{Luiijf2015}
A.~K. Eric~Luiijf.
\newblock Sharing cyber security information.
\newblock (March), 2015.

\bibitem{eu2016nisdirective}
C.~o. t. E.~U. European~Parliament.
\newblock Directive of the european parliament and of the council concerning
  measures for a high common level of security of network and information
  systems across the union, 2016.
\newblock PE 26 2016 INIT - 2013/027 (OLP).

\bibitem{freudiger2015controlled}
J.~Freudiger, E.~De~Cristofaro, and A.~E. Brito.
\newblock Controlled data sharing for collaborative predictive blacklisting.
\newblock In {\em Detection of Intrusions and Malware, and Vulnerability
  Assessment}, pages 327--349. Springer, 2015.

\bibitem{gambetta2000can}
D.~Gambetta et~al.
\newblock Can we trust trust.
\newblock {\em Trust: Making and breaking cooperative relations}, 13:213--237,
  2000.

\bibitem{garvey2009introduction}
P.~R. Garvey and C.~A. Pinto.
\newblock Introduction to functional dependency network analysis.
\newblock In {\em The MITRE Corporation and Old Dominion, Second International
  Symposium on Engineering Systems, MIT, Cambridge, Massachusetts}, 2009.

\bibitem{granatyr2015trust}
J.~Granatyr, V.~Botelho, O.~R. Lessing, E.~E. Scalabrin, J.-P. Barth{\`e}s, and
  F.~Enembreck.
\newblock Trust and reputation models for multiagent systems.
\newblock {\em ACM Computing Surveys}, 48(2):27, 2015.

\bibitem{guariniello2014communications}
C.~Guariniello and D.~DeLaurentis.
\newblock Communications, information, and cyber security in
  systems-of-systems: Assessing the impact of attacks through interdependency
  analysis.
\newblock {\em Procedia Computer Science}, 28:720--727, 2014.

\bibitem{hernandez2013information}
J.~L. Hernandez-Ardieta, J.~E. Tapiador, and G.~Suarez-Tangil.
\newblock Information sharing models for cooperative cyber defence.
\newblock In {\em Cyber Conflict (CyCon), 2013 5th International Conference
  on}, pages 1--28. IEEE, 2013.

\bibitem{eeuu2008nspd54}
T.~W. House.
\newblock National security presidential directive/nspd-54. homeland security
  presidential directive/hspd-23, 2008.
\newblock NSPD-54/HSPD-23.

\bibitem{ponemon2016costs}
P.~Institue.
\newblock 2016 cost of data breach study: global analysis.
\newblock Technical report, Ponemon Institute, 2016.

\bibitem{izquierdo2012learning}
L.~R. Izquierdo, S.~S. Izquierdo, and F.~Vega-Redondo.
\newblock Learning and evolutionary game theory.
\newblock In {\em Encyclopedia of the Sciences of Learning}, pages 1782--1788.
  Springer, 2012.

\bibitem{karsberg2015report}
C.~Karsberg and C.~Skouloudi.
\newblock Annual incident reports 2014.
\newblock Technical report, ENISA, 2015.

\bibitem{khouzani2014strategic}
M.~Khouzani, V.~Pham, and C.~Cid.
\newblock Strategic discovery and sharing of vulnerabilities in competitive
  environments.
\newblock In {\em Decision and game theory for security}, pages 59--78.
  Springer, 2014.

\bibitem{laube2015mandatory}
S.~Laube and R.~B{\"o}hme.
\newblock Mandatory security information sharing with authorities: Implications
  on investments in internal controls.
\newblock In {\em Proceedings of the 2nd ACM Workshop on Information Sharing
  and Collaborative Security}, pages 31--42. ACM, 2015.

\bibitem{li2007design}
J.~Li.
\newblock How to design a strategy to win an ipd tournament.
\newblock {\em The iterated prisoner’s dilemma}, 20:89--104, 2007.

\bibitem{meng2015collaborative}
G.~Meng, Y.~Liu, J.~Zhang, A.~Pokluda, and R.~Boutaba.
\newblock Collaborative security: A survey and taxonomy.
\newblock {\em ACM Computing Surveys (CSUR)}, 48(1):1, 2015.

\bibitem{naghizadeh2016inter}
P.~Naghizadeh and M.~Liu.
\newblock Inter-temporal incentives in security information sharing agreements.
\newblock In {\em Position paper for the AAAI Workshop on Artificial
  Intelligence for Cyber-Security}, 2016.

\bibitem{oliva2010agent}
G.~Oliva, S.~Panzieri, and R.~Setola.
\newblock Agent-based input--output interdependency model.
\newblock {\em International Journal of Critical Infrastructure Protection},
  3(2):76--82, 2010.

\bibitem{petrenj2013information}
B.~Petrenj, E.~Lettieri, and P.~Trucco.
\newblock Information sharing and collaboration for critical infrastructure
  resilience--a comprehensive review on barriers and emerging capabilities.
\newblock {\em International Journal of Critical Infrastructures},
  9(4):304--329, 2013.

\bibitem{rutkowski2010cybex}
A.~Rutkowski, Y.~Kadobayashi, I.~Furey, D.~Rajnovic, R.~Martin, T.~Takahashi,
  C.~Schultz, G.~Reid, G.~Schudel, M.~Hird, et~al.
\newblock Cybex: The cybersecurity information exchange framework (x. 1500).
\newblock {\em ACM SIGCOMM Computer Communication Review}, 40(5):59--64, 2010.

\bibitem{payoffdef}
M.~Shor.
\newblock "payoff," dictionary of game theory terms.
\newblock \url{http://www.gametheory.net/dictionary/Payoff.html}.
\newblock Accessed: 2016-07-11.

\bibitem{Spijkervet2014}
L.~B. Spijkervet.
\newblock Less is more.
\newblock Master's thesis, Delft University of Technology, 2014.

\bibitem{subramanian2014study}
S.~Subramanian and D.~e.~a. Robinson.
\newblock 2014 deloitte-nascio cybersecurity study. state governments at risk:
  time to move forward.
\newblock Technical report, Deloitte, NASCIO, 2014.

\bibitem{tosh2015evolutionary}
D.~Tosh, S.~Sengupta, C.~Kamhoua, K.~Kwiat, and A.~Martin.
\newblock An evolutionary game-theoretic framework for cyber-threat information
  sharing.
\newblock In {\em IEEE International Conference on Communications}, pages
  7341--7346. IEEE, 2015.

\bibitem{wang2013asymmetric}
Y.~WANG and C.~Ng.
\newblock Asymmetric payoff mechanism and information effects in water sharing
  interactions: A game theoretic model of collective.
\newblock In {\em International Komosozu Society, Mt. Fuji, Japan, 2013},
  page~68. IASC, 2013.

\end{thebibliography}
%
%

\end{document}